\documentclass[amsmath,amssymb]{article}

\usepackage{graphicx}
\usepackage{dcolumn}
\usepackage{bm}
\usepackage{wrapfig}
\usepackage{graphicx}
\usepackage{color}
\usepackage[latin1]{inputenc}
\usepackage{graphicx}
\usepackage{graphics}
\usepackage{amsmath}
\usepackage{amssymb}
\usepackage{amsfonts}
\usepackage{mathrsfs}
\usepackage{stmaryrd}
\usepackage{latexsym}

\renewcommand{\theequation}{\arabic{section}.\arabic{equation}}

\newcommand\bb{\textbf{\emph{b}}}

\newcommand\be{\textbf{\emph{e}}}

\newcommand\bff{\textbf{\emph{f}}}

\newcommand\bx{\textbf{\emph{x}}}

\newcommand\bs{\textbf{\emph{s}}}
\newcommand\bn{\textbf{\emph{n}}}

\newcommand\f{\textbf{\emph{f}}}

\newcommand\bF{\textbf{\emph{F}}}

\newcommand\bD{\textbf{\emph{D}}}

\newcommand\bE{\textbf{\emph{E}}}
\newcommand\bX{\textbf{\emph{X}}}

\newcommand\0{\textbf{\emph{0}}}

\newcommand\F{\textbf{F}}

\newcommand\B{\textbf{B}}

\renewcommand\S{\textbf{S}}

\newcommand\I{\textbf{I}}
\newcommand\T{\textbf{T}}

\renewcommand\d\delta
\newcommand\D\Delta

\newcommand\grad{\text{grad}}

\renewcommand\div{\text{div}}

\newcommand\e{\varepsilon}

\newcommand\scrE{\mathscr{E}}

\newcommand{\kv}{\kappa_{\mbox{\tiny \it V}}}
\newcommand{\kq}{\kappa_{\mbox{\tiny \it Q}}}
\newcommand{\kvp}{\kappa_{\mbox{\tiny \it V}}^{\mbox{\tiny \it pull}}}
\newcommand{\kvc}{\kappa_{\mbox{\tiny \it V}}^{\mbox{\tiny \it comp}}}

\begin{document}

\title{Failure mechanisms in thin electroactive polymer actuators}

\author{D. De Tommasi, G. Puglisi, G. Saccomandi, G. Zurlo}

\maketitle

\section*{Abstract} \begin{small}\it{ We propose a model to analyze the insurgence of pull-in and
wrinkling failures in electroactive thin films. We take in consideration both
cases of voltage and charge control, and study the role of prestretch and size
of activated regions, which are essential in the analysis of realistic
applications of EAPs. Based on simple geometrical and material assumptions we
deduce an explicit analytical description of these phenomena, allowing a clear
physical interpretation. Despite our simplifying assumptions, the comparison
with experiments shows a satisfying qualitative and, interestingly,
quantitative agreement. In particular our model shows, in accordance with
experiments, the existence of different optimal prestretch values, depending on
the choice of the actuating parameter of the EAP.}
\end{small}

\section{Introduction}

\noindent Electro-active polymers (EAPs) are very promising materials for
several technological applications in the robotic, medical, and biological
fields \cite{KP, CRK}. These materials are characterized by important qualities
such as lightweight, small size, low-cost, flexibility, fast response. The
rapid technological advances in material science have recently improved EAPs
performances in terms of their traditional limits, such as small actuation
forces, low robustness and the requirement of very high electric fields. A
prototypical device for electro-actuation based on EAPs is constituted by a
{\it thin} layer of polymeric material sandwiched between two compliant
electrodes. Application of a voltage to the two parallel electrodes generates
an electric force (measured through the so called Maxwell stress \cite{DO})
which induces a compression of the layer; since the polymer is typically nearly
incompressible, the resulting transversal extension is used as a mean of
actuation.

A typical drawback in technological applications based on the
described electro-activated films is the insurgence of different
types of instability, leading to failure (e.g. \cite{PD}). One of
the well known phenomena occurring in these devices is the so
called \textit{pull-in instability}. In order to describe this
type of instability, one may consider a toy model composed by two
rigid conducting plates connected by an insulated linear spring,
with a rest distance equal to $d$. If the plates are subjected to
a voltage $V$, the Coulomb forces between the plates will tend to
attract them so that the spring will result compressed. It is easy
to check that there exists a voltage threshold $V^*$ such that for
larger voltages no more equilibrium is possible between the
elastic and electric forces and the top plate slams onto the
bottom plate. The resulting \emph{thumb rule} used in EAPs and
similar MEMS devices is that the critical distance associated with
$V^*$ equals $d/3$.

Clearly, the quantitative determination of the pull-in voltage is
not a simple task for real deformable devices: the failure
mechanism of electro-activated thin films depends not only on
electro-elastic interactions, but also on the actuator shape and
on purely electrical effects. A further purely mechanical failure
mechanism in EAPs is represented by the insurgence of compressive
stresses inside the polymer film. In typical actuation systems
(e.g.\cite{KP},\cite{PD}) the polymeric film is constrained inside a
rigid frame which serves to assembly the actuator, to provide
electrical insulation, to transmit the actuation force, and to
provide a pre-stretching of the polymer. Under the action of
electric forces the constrained polymer film may undergo in-plane
compressions, typically shed in evidence by the insurgence of a
buckling type instability known as {\it wrinkling} in membrane
theory (\cite{Pip, Ste}). Another
crucial aspect for technological applications is given by the observation that the
polymer film is tipically activated only
in an internal region, due again to both assembly and insulation
motivations.

In the recent past several theoretical and numerical analysis have
been proposed in order to elucidate the causes of the various
failure mechanisms of EAPs actuation devices.
The importance of prestraining has been clearly
evidenced in \cite{KP} and \cite{PKP}, where experimental
observations have shown the existence of optimal prestraining
values. An interesting analysis of the benefits of prestraining
has been carried on in \cite{K} for an unconstrained layer
activated on the whole surface; in the same work, the necessity of
a correct description of the strongly non linear material behavior
of the polymer has also been evidenced. A similar problem is
considered in \cite{PGS} where the authors have discussed the
relevance of an energetic analysis in the description of the
pull-in instability. A numerical analysis for a circular planar
actuator has been recently presented in \cite{PD}, where the main
role of prestretch and stretch rate up to failure are numerically
analyzed. An energetic analysis of prestress and pull-in for a
homogeneously deformed dielectric elastomers is studied in
\cite{ZS}. Finally, a theoretical analysis of the insurgence of deformation
localization in a variational framework was recently proposed in
\cite{GP} and \cite{ZHS}, where the role of damage and dissipation
were considered.

\begin{figure}[ht]\vspace{0.3cm}
\begin{center}
\includegraphics[height=6.5 cm]{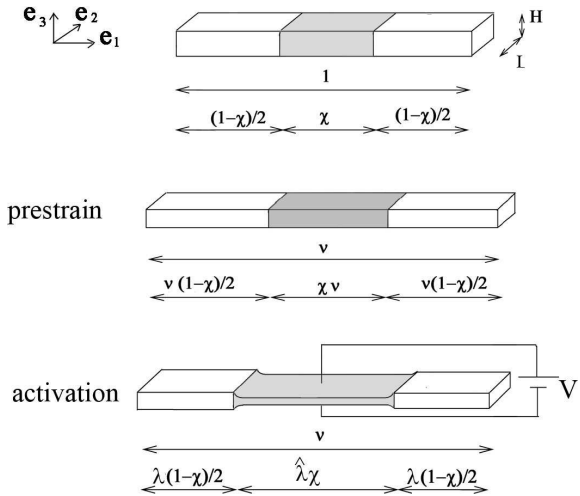}\vspace{0.3 cm} \caption{Scheme of the model and
boundary conditions. The gray region represents the activated
region. \label{figa}}
\end{center}
\end{figure}

In this paper we propose a simple prototypical model for a slender
prestrained actuator with an inner region activated by assigned
voltage or charge (see the scheme in Fig.\ref{figa}). Based on
simplifying geometrical and material hypotheses we elucidate the
combined roles of prestrain, material properties and dimensions
with respect to the wrinkling and pull-in failure modes. Due to
the film thinness we neglect bending resistance and the formation
of a boundary layer between activated and non activated regions,
so that we assume a sharp interface between such regions
\cite{C, CGS} and we ignore fringing effects \cite{Par}. Our
geometrical assumptions are well suited for linear actuators
typically adopted in artificial muscle technologies \cite{PKP}.
Of course our approach can be extended to more complex
geometrical schemes and constitutive assumptions by using
standards numerical methods.

The paper is organized as follows. In section \ref{equi} we recall
the basic electromechanical equilibrium equations. In section
\ref{set} we introduce our prototypical example for a thin
electroactivated system. In section \ref{eqpr} we obtain
analytical solutions for both boundary value problems with
assigned charge and voltage, as a function of given prestretch.
Then, in section \ref{stabil}, we discuss the results of
our model. In particular, we deduce analytically the existence of
an optimal pre-strain (experimentally detected in \cite{PKP})
which depends on the material modulus of the polymeric layer and
on the size of the activated region. Shortly, the model describes
how prestretch has a positive effect on the wrinkling instability
and a negative effect on the pull-in instability, leading to the
existence of an optimal prestretch. The corresponding value sof the
activation parameters are then deduced. In section \ref{exp} we
show that the model well describes the experimental behavior.
Finally, in the Appendix we deduce a necessary stability condition
for the electroactive device here considered.
\section{Basic Equations}\label{equi}
In this section we collect the main equations that are at the base
of our theoretical approach. We refer the reader to the recent
paper \cite{DO} and to the references therein for details. Let
$\bX\in {\cal B}_{0}$ and $\bx \in \cal B$ be the typical  point
in the reference ${\cal B}_{0}$ and in the current configuration
$\cal B$ of a continuous body, with $\bx=\f(\bX )$, where $\f$ is
the deformation field. Moreover, we denote by $\phi$, $\bD$ and
$\bE$ the potential, the displacement vector and the electric
field in the current configuration, respectively. For a linear,
homogeneous and isotropic dielectric we have
\begin{equation} \bD=\e\bE=-\e \grad \phi,\label{const}\end{equation}
where $\e=\e_0\e_d$ with $\e_0$ the permittivity of free space and
$\e_d$ the dielectric constant of the material. The
equilibrium equations for the electromechanical problem are
\begin{equation}
\begin{array}{l}
\div \bD = \rho, \,\, \div \T = \bb \, \mbox{ in } \, \cal B \vspace{0.2 cm}\\
\llbracket\bD\rrbracket\cdot \bn = \hat q \, \mbox{ in } \,
\partial {\cal B}_{q}, \,\, \T\bn =\hat \bs \, \mbox{ in } \,
\partial {\cal B}_{s}.
\end{array}
\end{equation} Here $\rho$ and
$\hat q$ (assigned on $\partial {\cal B}_{q}$) are the bulk and
surface charge densities and $\bb$ and $\hat \bs$ (assigned on
$\partial {\cal B}_{s}$) are the bulk and surface forces densities
in the current configuration. $\T$ is the (symmetric) total stress
tensor, that can be decomposed as the sum of the elastic (ela)
term $\T^{{ela}}$ and of the electric (elc) component $\T^{{elc}}$
(i.e. the Maxwell stress)
\begin{equation}\label{abc}
\T=  \T^{ela}+ \T^{elc},\end{equation} with
\begin{equation}\begin{array}{c}
\T^{elc}=\e (\bE \otimes \bE-\frac{1}{2} (\bE \cdot \bE) \I ),\vspace{0.2 cm}\\
T^{elc}_{ij}=\frac{1}{2}E_{i}E_{j}- (\bE \cdot \bE) \delta_{ij},
\end{array}\label{def}
\end{equation}
where $i,j$ range from 1 to 3 (here and in the following we will
use, when necessary, both component and absolute notations). The
corresponding nominal (Piola-Kirchhoff) stress can be obtained as
\begin{equation}
\vspace{0.2 cm}\\  \S=  \S^{ela}+ \S^{elc}=J \T^{ela} \F^{-T}+J
\T^{elc} \F^{-T},\end{equation} where $\F:=\nabla \bff$ and
$J:=\det \F$.

\section{Setting of the problem} \setcounter{equation}{0}\label{set}

We consider a dielectric elastomer actuator which occupies, in its
natural configuration, a right prismatic region
$[0,\ell]\times[0,\tilde{L}]\times[0,\tilde{H}]$. We use the
lenght $\ell$ of the elastomer to adimensionalize the thickness
$H=\tilde{H}/\ell$ and the width $L=\tilde{L}/\ell$ and thus we
consider the prismatic region  $[0,1]\times[0,L]\times[0,H]$ (see
Fig.\ref{figa}, where we also indicate the chosen coordinate
system). In particular, we assume that the dielectric is
\emph{thin}, in the sense that $H\ll L\ll 1$. For thin films the
membrane approximation of the nonlinear elasticity may be adopted.
The standard \textit{membrane assumption} (e.g. \cite{HO}) states that
the bending stiffness is zero and any in-plane
compressive stress leads immediately to the membrane buckling,
with the appearance of a wrinkled regions. The presence of
wrinkles is frequently encountered in experiments on thin
dielectric elastomers and it represents one of the main observed
failure mechanism of such devices.

We assume that voltage or charge can be controlled on a region of
the thin film, more precisely on opposite strips of length
$\chi\leq 1$ belonging to the upper and lower faces of the
actuator. This region is placed centrally in direction of $\be_1$
length and it covers the full length $L$ in direction $\be_{2}$.
The prismatic region of the actuator with upper and lower faces
coinciding with the electrically controlled strips will be shortly
denoted by \emph{active} region, whereas the remaining part is
simply called \emph{non active} region. In accordance with the
technological applications described in the introduction, we
consider the possibility of a prestretch of the actuator along the
$\be_{{1}}$ direction (see again Fig.\ref{figa}).

We describe the layer deformation by two homogeneous
deformations with eigenvectors coinciding with the coordinate axes. More
specifically, we assume in the non active region a deformation
$\f$ such that
\begin{equation}\begin{array}{l}
\underline{ \mbox{non active region:}}\vspace{0.2 cm} \\
\begin{array}{c} x_{1}=\lambda_{1}X_{1}\\
x_{2}=\lambda_{2}X_{2}\\
x_{3}=\lambda_{3}X_{3}\end{array}, \hspace{0.3 cm}
\F:=\nabla\f=\left [
\begin{array}{ccc} \lambda_{1} & 0 & 0\\
0& \lambda_{2}& 0\\
0&0&\lambda_{3} \end{array}\right ].\end{array}
\label{eaa}\end{equation} In the activated region we
denote by $\hat{\bX}=(\hat X_{1},\hat X_{2},\hat X_{3})$ and
$\hat{\bx}=(\hat x_{1},\hat x_{2},\hat x_{3})$ the generic point
in the reference and current configuration, respectively, and by
$\hat{\f}$ the deformation. For this region we assume
\begin{equation}\begin{array}{l} \underline{ \mbox{active region:}} \vspace{0.2 cm}\\
\begin{array}{c} \hat x_{1}=\hat \lambda_{1}\hat X_{1}\\
\hat x_{2}=\hat \lambda_{2}\hat X_{2}\\
\hat x_{3}=\hat \lambda_{3}\hat X_{3}\end{array}, \hspace{0.3 cm}
\hat{\F}:=\nabla \hat{\f}=\left [
\begin{array}{ccc} \hat \lambda_{1} & 0 & 0\\
0& \hat \lambda_{2}& 0\\
0&0&\hat \lambda_{3} \end{array}\right ].\end{array}
\label{ebb}\end{equation}

The assumption of piecewise homogeneous deformation
can be justified by the following hypotheses: piecewise homogeneous
loading, thinness hypothesis, and constitutive assumption
of homogeneous, isotropic, material behavior of the polymer.
This type of deformations (which essentially neglects the presence of a
geometrically-compatibile boundary layer at the interface between
active and non active regions) is classical within the framework
of phase-transitions \cite{FU} and of thin bodies subjected to
homogeneous boundary loading (see, e.g.,
\cite{C, CGS}). The balance equation at the interface may be formulated
in the weak form $\llbracket\S\rrbracket\be_{1}=\0$, $\be_{1}$
being the normal to the interface in the reference configuration.

As well as the geometrical boundary layer we here neglect edge
effects of the electric field, known in literature as
\emph{fringing fields}. Across the interface between the active
and non active regions any discontinuity of the electric field
$\llbracket\bE\rrbracket$ must fulfill the jump condition (see
e.g. \cite{DO}) $ \be_{1}\times\llbracket\bE\rrbracket=\0$. The assumption of
negligible fringing fields leads to an apparent violation of the
jump condition, since it results that $\bE=E \be_{3}$ in the
active region whereas $\bE=\0$ in the non active region, so that the
component of $\bE$ which is parallel to the interface is not
continuous through the interface; this apparent violation is
actually resolved by considering that in reality the electric
field undergoes an abrupt, but continuous, variation from $E$ to
$0$ in a region of negligible width. Also the assumption regarding
electrical edge effects are acceptable,
granted the thickness of the EAP is much smaller than its other
dimensions (see e.g. \cite{Par}).

Regarding the polymer material
behavior we assume that this is incompressible, so that its
deformations respect the isochoric constraint
\begin{equation}\label{incompr}
\hat{\lambda}_1\hat{\lambda}_2\hat{\lambda}_3=1,\hspace{30pt}\lambda_1\lambda_2\lambda_3=1.
\end{equation}

Moreover, in a typical application scheme, the actuator is
pre-stretched in the direction $X_1$, say of an amount equal to
$\nu$. By imposing that the total length of the actuator in the
current configuration equals $\nu$, we get
\begin{equation}\label{elongation}
\nu = \chi\hat{\lambda}_1 + (1-\chi)\lambda_1.
\end{equation}
To obtain analytic results we assume that the incompressible
polymer is characterized by a neo-Hookean constitutive response
\begin{equation}
\begin{array}{l}
\T^{ela} = -\tilde \pi \I + \mu \B, \,\,\,\, \S^{ela}=\T^{ela} \F^{-T}, \,\,\,\,  \\
T^{ela}_{ij}=-\tilde \pi \delta_{ij}+\mu \lambda_{i}^{2}, \,\,\,\,
S^{ela}_{ij}=-\tilde \pi \lambda_{i}^{-1}\delta_{ij}+\mu
\lambda_{i},
\end{array}
\end{equation}
where $\tilde \pi$ is a Lagrange multiplier due to the
incompressibility constraint, $\mu$ is the shear
modulus and $\B:=\F \F^{T}$ is the left Cauchy-Green deformation
tensor. By introducing the adimensionalized principal stresses and
pressure
$$
s_{i}:=\frac{S_{i}}{\mu},\hspace{10pt}t_{i}:=\frac{T_{i}}{\mu},\hspace{10pt}
\pi:=\frac{\tilde \pi}{\mu},
$$
the non zero stress components in the non active region can be
calculated by substitution of the deformation (\ref{eaa}) in the
constitutive response, which gives
\begin{equation}\label{stress general}
\begin{array}{l}
\underline{ \mbox{non active region:}}\vspace{0.2 cm} \\
s_i = -\pi\lambda_i^{-1} + \lambda_i,\hspace{10pt} t_i = -\pi +
\lambda_i^2\hspace{10pt}(i=1,2,3).
\end{array}
\end{equation}

As anticipated above, we assume that the electric field in the active region coincides
with $\bE=(0,0,E)$. By using (\ref{abc}) and (\ref{def}) the
principal stresses in the active region are given by
\begin{equation}\begin{array}{l}
\underline{ \mbox{active region:}}\vspace{0.2 cm} \\
\hat{t}_i=-\pi+
\hat\lambda_i^2+\e\frac{E^2}{\mu}\left(\delta_{i3}-\frac{1}{2}\right),
\\ \hat s_{i}=-\pi \hat \lambda_{i}^{-1}+
\hat\lambda_i+\e\frac{E^2}{\mu}\left(\delta_{i3}-\frac{1}{2}\right)
\hat\lambda_{i}^{-1}. \end{array} \label{Maxwell 1}
\end{equation}

The nature of the electric stresses, as well as their geometrical
coupling with the deformation, changes according to the fact that
\emph{voltage} or \emph{electric charge} are externally controlled
by the device which drives the actuator. In the first
case, the applied voltage $V$ determines the electric field:
\begin{equation}\label{VOLT}
\mbox{\underline{voltage control}: }\bD\!=\!(0,0,\!\frac{\e
V}{H\hat{\lambda}_3}\!),
\bE\,=\!(0,0,\frac{V}{H\hat{\lambda}_3}\!).
\end{equation}
In the second case, the total electric charge $Q$ is controlled on
the electrodes, so that
\begin{equation}\label{CHARGE}
\mbox{\underline{charge
control}}\!:\!\bD\!=\!(0,\!0,\!\!\frac{Q}{\hat\lambda_1\hat
\lambda_2 A}\!), \bE\!=\!(0,\!0,\!\!\frac{Q}{\e \hat\lambda_1 \hat
\lambda_2 A}\!).\!\!
\end{equation}
Here $A=\chi\,L$ is the reference area where the charge is
applied. In the following the two driving mechanisms will
be considered separately.

\section{Equilibrium for the voltage and charge controls}\setcounter{equation}{0}\label{eqpr}

We first consider the following  pre-stretch deformation $\f^{p}$
in the $\be_1$ direction
\begin{equation}
\begin{array}{c} x^{p}_{1}=\nu X_{1}\\
x^{p}_{2}=\nu_{2}X_{2}\\
x^{p}_{3}=\nu_{3}X_{3}\end{array}, \hspace{0.5 cm}
\bF^{p}:=\nabla_{\bX}\f^{p}=\left [
\begin{array}{ccc} \nu & 0 & 0\\
0& \nu_{2}& 0\\
0&0&\nu_{3} \end{array}\right ],
\end{equation}
where $\nu=(\nu_2\nu_3)^{-1}$ is the assigned longitudinal
prestretch. (Here and in the following, for simplicity of notation,
we sometime drop the index $1$ when there is no ambiguity.)

The principal Cauchy stresses in the prestretched configuration
are given by
$$
t^{p}_i=-\pi^{p}+\nu_i^2,
$$
so that after imposing the conditions of vanishing stresses on the
faces orthogonal to the axis $X_2$ and $X_3$
\begin{equation}\label{free}
t^{p}_2=t^{p}_3=0,
\end{equation}
we obtain $\pi^{p}=\nu^{-1}$ and $\nu_2=\nu_3=\nu^{-1/2}$. Thus,
the only non-zero stress component in the prestretched
configuration amounts to
\begin{equation}\label{stress P}
t^{p}_1=\nu^2-\nu^{-1}, \hspace{30pt}s^{p}_1=\nu-\nu^{-2}.
\end{equation}

Now we consider the electric activation of the EAP.
Regarding the nonactive region the boundary conditions $t_{2}=t_{3}=0$ give
$(\lambda_1,\lambda_2,\lambda_3)=(\lambda,\lambda^{-1/2},\lambda^{-1/2})$
so that the stress in this region as in
Eq.ns (\ref{stress P}) is given by
\begin{equation}\label{Stress a}
t_i=(\lambda^2-\lambda^{-1})\delta_{i1},\hspace{20pt}s_i=(\lambda-\lambda^{-2})\delta_{i1}.
\end{equation}
Concerning the electrically active region, after substitution of
Eq.ns (\ref{VOLT}) in Eq.(\ref{Maxwell 1}), we get
\begin{equation}\label{TbV}
\mbox{\underline{voltage control}: }
\hat{t}_i=-\hat{\pi}+\hat{\lambda}_i^2+\frac{1}{2
\mu}\frac{\e V^2}{\hat{\lambda}_3^2H^2}(2\delta_{i3}-1),
\end{equation}
and, by using (\ref{incompr}), (\ref{Maxwell 1}) and
(\ref{CHARGE}) we get
\begin{equation}\label{TbQ}
\mbox{\underline{charge control}: } \hat{t}_i=-\hat{\pi}+\hat
\lambda^2_i+\frac{1}{2\mu}\frac{Q^2\hat{\lambda}_3^2}{\e
A^2}(2\delta_{i3}-1).
\end{equation}

In order that the EAP maintains its total length equal to the
given prestretch also after the application of charge or voltage (see Eq.(\ref{elongation})), we impose
\begin{equation}\nu=(1-\chi)\lambda+\chi\hat{\lambda}.\label{compat}\end{equation}
This equation gives the following relation
\begin{equation}\label{lambda}
\lambda=\frac{\nu-\chi\hat{\lambda}}{1-\chi},
\end{equation}
so that the condition $\lambda>0$ implies that
\begin{equation}\hat{\lambda}<\nu/\chi
\label{ddt}\end{equation}
Using (\ref{Stress a}) and (\ref{lambda}) we then obtain
\begin{equation}\label{Sa11}
s=\left(\frac{\nu-\chi\hat{\lambda}}{1-\chi}\right)-
\left(\frac{\nu-\chi\hat{\lambda}}{1-\chi}\right)^{-2}.
\end{equation}

Let us now take in consideration the case when voltage is
applied. In this case it is convenient to introduce the parameter
$\kv$ defined by
$$
\kv=\frac{1}{2\mu}\frac{\e V^2}{H^2},
$$
which represents the (adimensionalized) electric free energy
density under the assigned potential. By imposing the boundary
conditions $\hat{t}_{2}=\hat{t}_{3}=0$, after simple calculations
we obtain the equilibrium stress and strains
\begin{equation}\label{Sb11} \mbox{\underline{voltage control}: }\left \{ \begin{array}{l}\displaystyle
\hat{s}=\hat{\lambda}-\frac{1}{\hat{\lambda}^2\sqrt{1-2\kappa_{\mbox{\it
\tiny V}}\hat{\lambda}^2}},\vspace{0.2 cm}\\ \displaystyle \hat
\lambda_{2}^2=\frac{1}{\hat \lambda \sqrt{1-2\kappa_{\mbox{\it
\tiny V}}\hat{\lambda}^2}},\vspace{0.2 cm}\\ \displaystyle \hat
\lambda_{3}^2=\frac{\sqrt{1-2\kappa_{\mbox{\it \tiny
V}}\hat{\lambda}^2}}{\hat \lambda}.\end{array}\right .
\end{equation}
When the total charge is assigned, after imposing the boundary
conditions $\hat{t}_{2}=\hat{t}_{3}=0$ and by introducing the
(adimensionalized) electric energy density $\kappa_{\mbox{\tiny
\it Q}}$
$$
\kappa_{\mbox{\tiny \it Q}}=\frac{1}{2\mu}\frac{Q^2}{\e A^2},
$$
we obtain
\begin{equation}\label{Sc11}
\mbox{\underline{charge control}: }\left \{ \begin{array}{l}\displaystyle
\hat{s}=\hat{\lambda}-\frac{\sqrt{1+2\kappa_{\mbox{\tiny{\it
Q}}}}}{\hat{\lambda}^2},\vspace{0.2 cm}\\ \displaystyle
\lambda_{2}^2=\frac{\sqrt{1+2\kappa_{\mbox{\tiny \it Q}}}}{\hat
\lambda},\vspace{0.2 cm}\\
\displaystyle\lambda_{3}^2=\frac{1}{\hat
\lambda\sqrt{1+2\kappa_{\mbox{\tiny \it Q}}}}.\end{array} \right .
\end{equation}

Finally, in order to deduce the equilibrium configurations we must impose
the mechanical balance at the interface between the active and
nonactive regions, which reads
\begin{equation}\label{equilibrium}
s = \hat{s}.
\end{equation}
This equation let us determine $\hat{\lambda}$ and, hence, the
equilibrium configurations of the system.

\section{Wrinkling and pull-in instabilities}\setcounter{equation}{0}\label{stabil}

The two failure mechanisms of wrinkling and pull-in will be
considered separately for the cases of voltage and charge control.

\subsection{Voltage Control}
\begin{figure}[h!]
\vspace{0 cm}
\begin{center}
\hspace{-0.55 cm}\includegraphics[height=6.15 cm]{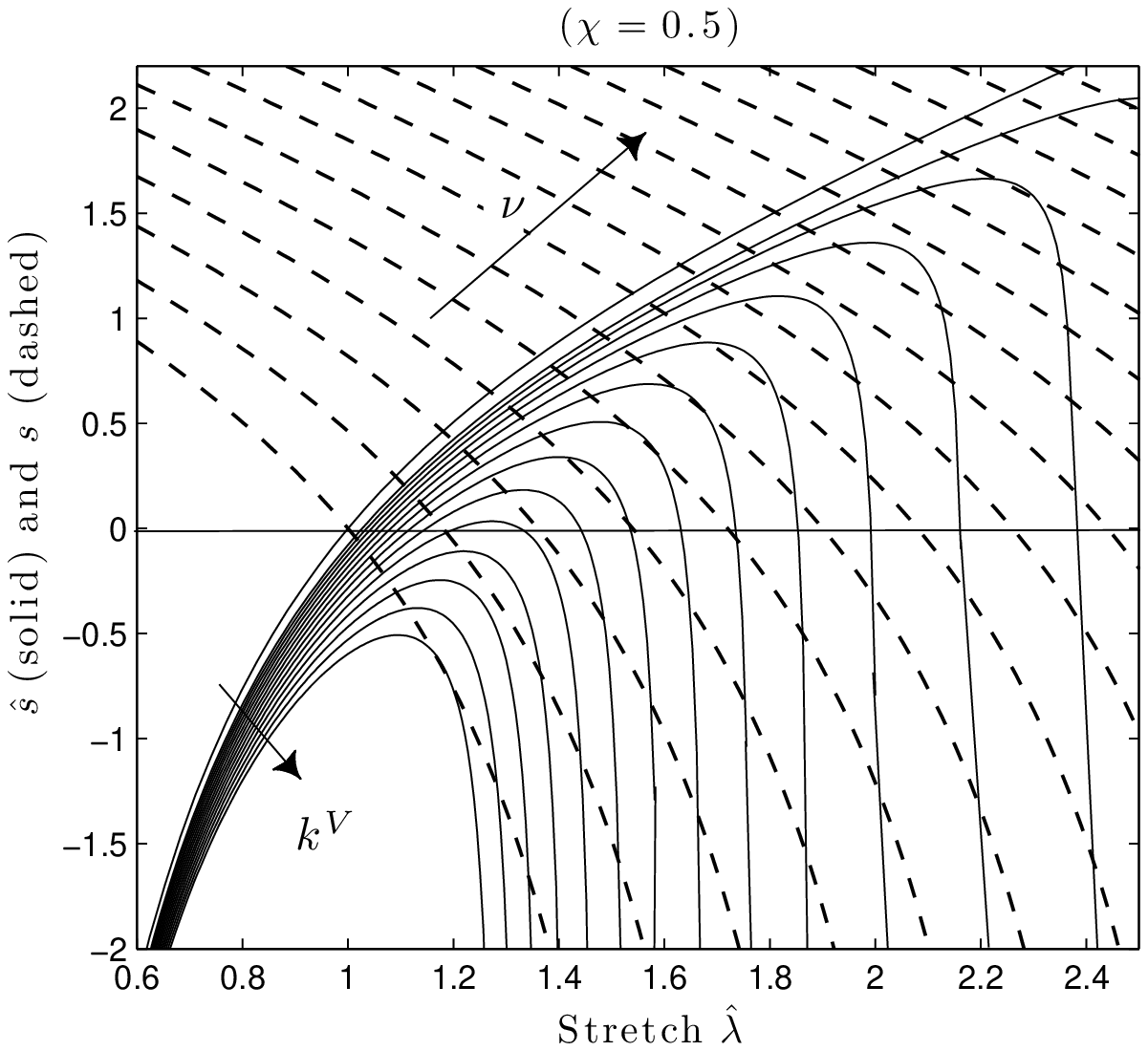}\vspace{0.5 cm}a)\\
\hspace{0.1 cm}\includegraphics[height=5.7 cm]{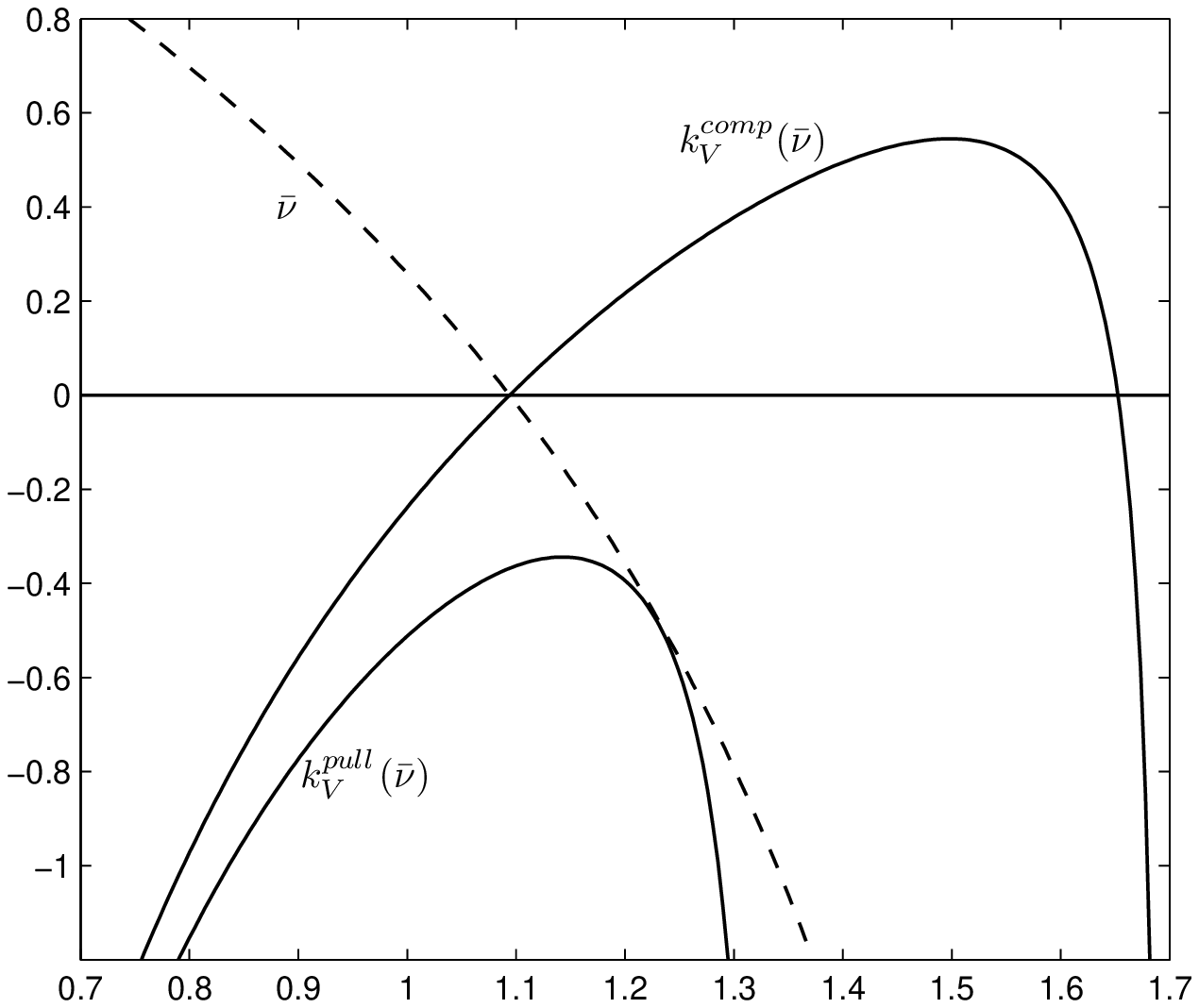}b)
\hspace{.9 cm}\includegraphics[height=5.4 cm]{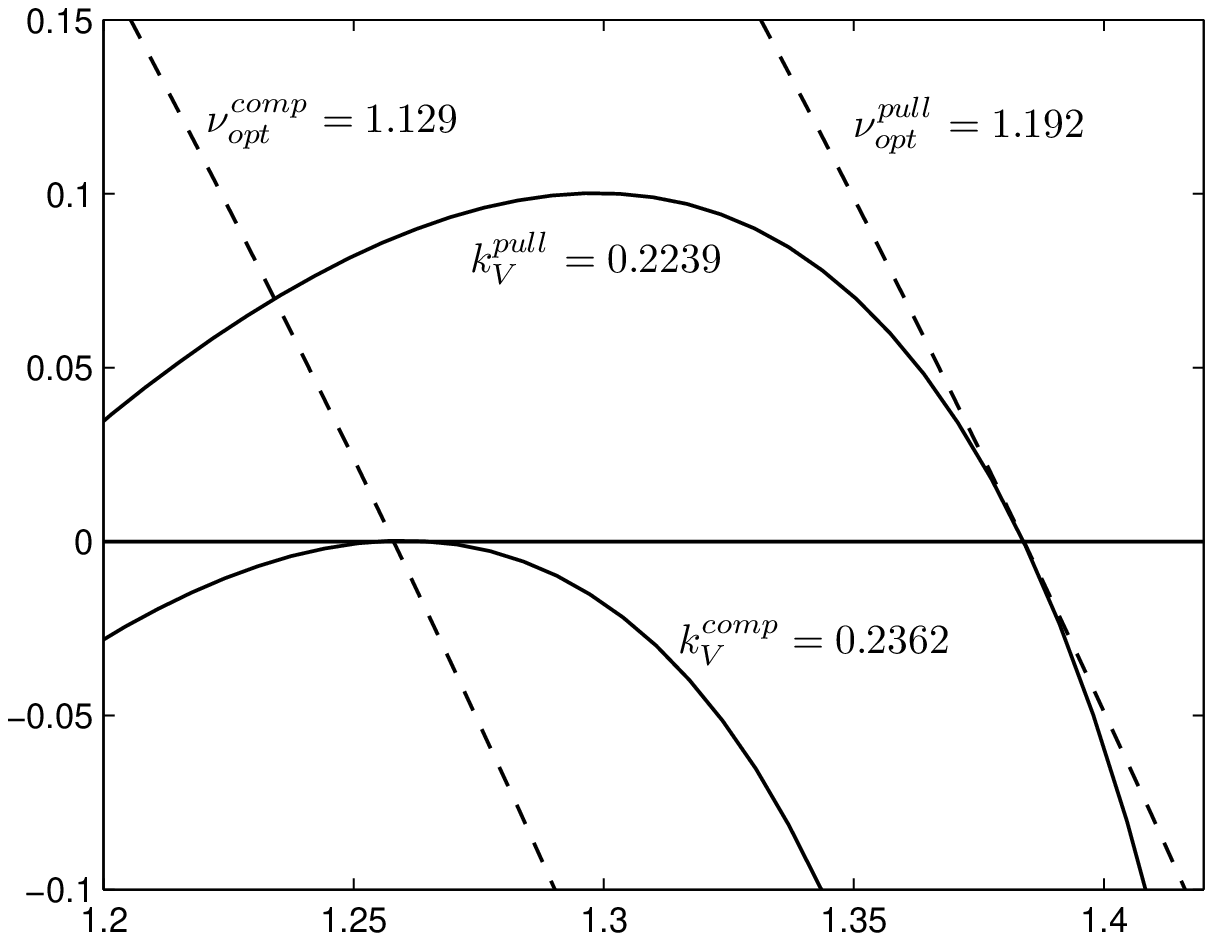}c)\\
\vspace{.6 cm}\caption{\label{Eqvolt} Stress strain curves in the activated and
non activated region in the case of assigned voltage: continuous lines
represent the stress $\hat{s}_{1}(\hat \lambda, \kv)$ in (\ref{Sb11}), of the
activated region, whereas dashed lines represent the stress $s_{1}(\hat
\lambda, \nu)$ in (\ref{Sa11}) in the non active region. In b) we show the
attainment, for a fixed $\nu$ of both the pull-in and wrinkling instability. In
c) we show the stress curves $s_{1}(\hat \lambda, \nu)$ corresponding to the
maximum attained value of $\kappa_{\mbox \tiny{V}}$ }
\end{center}
\end{figure}

When voltage is assigned (see Fig.\ref{Eqvolt}) the equilibrium
configurations, according with (\ref{equilibrium}), (\ref{Sa11}), (\ref{Sb11}) correspond (if these exist)
to the zeros of the function
$$
g(\hat{\lambda})\equiv
\underbrace{\hat{\lambda}-\frac{1}{\hat{\lambda}^2\sqrt{1-2\kappa_{\mbox{\it
\tiny
V}}\hat{\lambda}^2}}}_{\hat{s}}-\underbrace{\left[\frac{\left(\frac{\nu-\chi
\hat{\lambda}}{1-\chi}\right)^3-1}{\left(\frac{\nu-\chi\hat{\lambda}}{1-\chi}\right)^{2}}\right]}_{s},
$$
measuring the stress discontinuity at the interface. In order to obtain physically reasonable deformations (see (\ref{ddt}) and (\ref{Sb11})$_{1}$) we are
interested to the solutions of $g(\hat{\lambda})=0$ in the range
$(0, \hat{\lambda}^*)$ where $\hat{\lambda}^*$ is the
$\min\{1/\sqrt{2\kv}, \nu/\chi\}$. We observe that the stress
$s$ in (\ref{Sa11}) is a ($\nu$--dependent)
monotonic decreasing function of $\hat \lambda$, whereas an
analysis of (\ref{Sb11}) shows that $\hat s$ is a
($\kv\,$--dependent)  concave function of $\hat \lambda$, defined
for $\hat \lambda \in (0,1/\sqrt{2\kv})$, with $\hat
s\rightarrow -\infty$ at the boundaries of the definition
domain. An inspection of the stress plots reported in
Fig.\ref{Eqvolt}$_{a}$ shows that for given $\nu$ and $\chi$ the
condition $g(\hat{\lambda})=0$ may have two roots
$\hat{\lambda}_{-} \leq \hat{\lambda}_+$ (equal or distinct) or no
roots in the range of interest. If two roots exist it is
possible to show that only one of them is \emph{stable} in a sense
which is discussed in detail in the Appendix; by means of this
condition, it easy to check that the stable solution corresponds
to the root $\hat{\lambda}_{-}$.

For given $\chi,\nu$ we define
$\kvc$ the value of the loading parameter in correspondence of
which the stress evaluated at $\hat\lambda_-$ is such that
$s=\hat s=0$ (see Fig.\ref{Eqvolt}$_{b}$): for values of $\kv>\kvc$ the equilibrium stress
is negative and, due to the lack of bending resistance, the
film exhibits wrinkling, determining the device failure.
Always keeping $\nu,\chi$ fixed, a second eventuality corresponds
to values of the load parameter for which the two solutions
coalesce  (see again Fig.\ref{Eqvolt}$_{b}$), so that for values of $\kv$ larger than this critical
value $\kvp$ no more roots of $s=\hat s$ exist: this instance
represents the pull-in instability. For $\kv>\kvp$ the elastic
response of the film is no more capable of balancing the
compressive electric forces and the facing electrodes smash onto
each other. Clearly, in correspondence of $\kv=\kvp$ it results
$g=0$ and $dg/d\hat{\lambda}=0$.

\begin{figure}[t!]
\begin{center}\includegraphics[height=6.7 cm, width=8 cm]{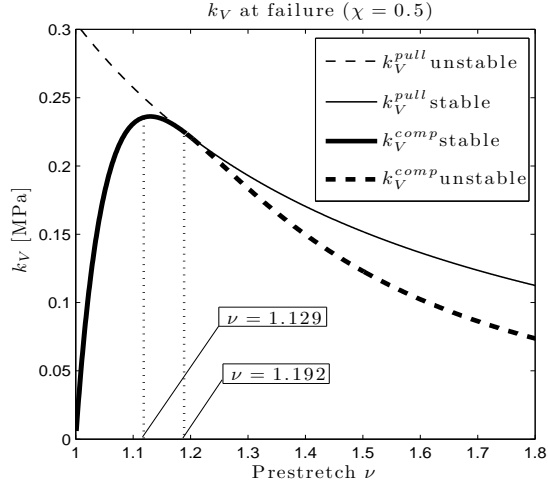}
 \vspace{0
cm}\caption{\label{optV} Critical values of the activation energy
$k_{\mbox{\it \tiny V}}$. Bold line represents the function $k_{\mbox{\tiny \it V}}^{\mbox{\tiny \it comp}}(\nu)$ corresponding to wrinkling instability, whereas light line represent the function  $k_{\mbox{\tiny \it V}}^{\mbox{\tiny \it pull}}(\nu)$ corresponding to pull-in instability. Dashed lines correspond to unstable configurations.}
\end{center}
\end{figure}

The situation is depicted in Fig.\ref{optV}, where
both values of $\kvc$ and of $\kvp$ are plotted with respect to
the prestretch $\nu$. Under the particularly simple assumptions of
our model, the expression of $\kvc$ can be deduced analytically by
imposing $s_1=\hat s_1=0$, which gives
$$
\kvc(\nu)=\frac{1}{2} \left(\frac{1}{\hat
\lambda^{2}(\nu)}-\frac{1}{\hat \lambda^{8} (\nu)}\right), \mbox{
with } \hat \lambda(\nu)=\frac{\nu+\chi-1}{\chi}.
$$
The function $\kvc(\nu)$ is plotted with a bold line in
Fig.\ref{optV}; it attains a maximum in correspondence of a
prestretch
\begin{equation}
\nu_{\mbox{\tiny \it opt}}^{\mbox{\tiny \it comp}}=1+({2}^{1/3}
-1)\chi.\label{nioptw}
\end{equation}
The maximum value of $\kv$ which can be applied to the film
without inducing compression then amounts to
\begin{equation}\kappa_{\mbox{\tiny \it V,max}}^{\mbox{\tiny \it comp}}=\frac{1}{2}(2^{-2/3}-2^{-8/3}),\label{koptw}\end{equation}
attained when
$$\hat \lambda=\hat \lambda_{\mbox{\tiny \it opt}}^{\mbox{\tiny \it comp}}=2^{1/3}.$$

The analytical expression of $\kvp$ is not easy to get, but the
numerical solution of the conditions $g=0$, $dg/d\hat{\lambda}=0$
gives $\kvp$ for each value of $\nu$, which is represented by a
light line in Fig.\ref{optV}.
It is possible to show that, for fixed
$\chi$, the curve $\kvc(\nu)$ is always under the curve
$\kvp(\nu)$ but in a common point where these curves are tangent,
corresponding to the simultaneous attainment of both wrinkling and
pull-in failures. This point can be analytically determined by
imposing that $s=\hat s=0$ and $dg/d\hat{\lambda}=0$, which
after simple algebraic manipulations yields the value of the
prestretch
\begin{equation}
\nu_{\mbox{\tiny \it opt}}^{\mbox{\tiny \it pull}}=\chi(\hat
\lambda_{\mbox{\tiny \it opt}}^{\mbox{\tiny \it pull}}(\chi)-1)+1,
\label{nipi}
\end{equation}
where
\begin{equation}\label{chipi}
\hat \lambda_{\mbox{\tiny \it opt}}^{\mbox{\tiny \it pull}}(\chi):
= \left(\frac{4-\chi}{1-\chi}\right)^{1/6}.
\end{equation}

\noindent It is easy to check that for all $\chi$ it results
$\nu_{\mbox{\tiny \it opt}}^{\mbox{\tiny \it
comp}}\leq\nu_{\mbox{\tiny \it opt}}^{\mbox{\tiny \it pull}}$.

It
is interesting to observe that not all the values of the functions
$\kvp(\nu)$ and $\kvc(\nu)$ are actually attainable; regarding
$\kvp(\nu)$, the analysis of the functions $s(\hat \lambda)$ and
$\hat s(\hat \lambda)$ shows that for $\nu>\nu_{\mbox{\tiny \it
opt}}^{\mbox{\tiny \it pull}}$ the pull-in failure is attained at
positive stress, whereas  for $\nu<\nu_{\mbox{\tiny \it
opt}}^{\mbox{\tiny \it pull}}$ it is attained at negative stress,
which is not possible in our model (dashed light line in
Fig.\ref{optV}). On the other side it is possible to check that
the values of $\kvc$ corresponding to $\nu>\nu_{\mbox{\tiny \it
opt}}^{\mbox{\tiny \it pull}}$ are relative to the unstable
solutions $\hat\lambda_+$ and hence these are not attainable
(dashed bold line in Fig.\ref{optV}). As a result, the film
failure is due to wrinkling for $\nu<\nu_{\mbox{\tiny \it
opt}}^{\mbox{\tiny \it pull}}$ and it is due to pull-in for
$\nu>\nu_{\mbox{\tiny \it opt}}^{\mbox{\tiny \it pull}}$. As it is
clear from the plot, the highest value of the load parameter which
can be applied without generating failure of both types coincides
with the maximum value of $\kvc$ given in Eq.\ref{koptw}, which is
attained in correspondence of $\nu=\nu_{\mbox{\tiny \it
opt}}^{\mbox{\tiny \it comp}}$; in this sense $\nu_{\mbox{\tiny
\it opt}}^{\mbox{\tiny \it comp}}$ is \emph{optimal} if one is
interested at the maximization of the stored electric energy.

\begin{figure}[th]
\begin{center}
\includegraphics[height=7 cm]{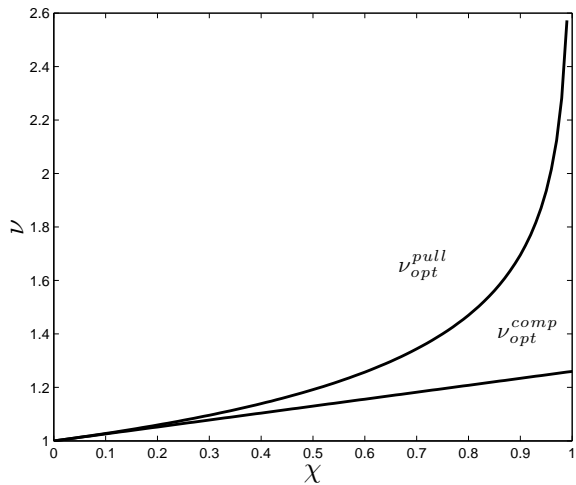} \vspace{0.3
cm}\caption{\label{optimni} Optimal prestretch (\ref{nioptw}) and (\ref{nipi})
as functions of the size  $\chi$ of the activation region.}
\end{center}
\end{figure}

A remarkable fact is that if one is interested at other useful
activation parameters, such as the maximum actuating force or the
maximum applicable electric field, then $\nu_{\mbox{\tiny \it
opt}}^{\mbox{\tiny \it comp}}$ is no more \emph{optimal}. This can
be seen by evaluating the values of the stress gap with respect to
the prestretched configuration, given by $\Delta S:=\mu[\hat
s(\nu)-\hat s(\hat \lambda)]$ and of the electric field $\bE$
in correspondence of the initiation of both wrinkling and pull-in
failures; as it is evident from the plots, the maximum values of
both parameters are now attained for $\nu=\nu_{\mbox{\tiny \it
opt}}^{\mbox{\tiny \it pull}}$, that is for $\hat \lambda=\hat
\lambda_{\mbox{\tiny \it opt}}^{\mbox{\tiny \it pull}}$ (see
Eqn.s(\ref{chipi})-(\ref{nipi})). We remind that this value of
$\nu$ corresponds to the simultaneous attainment of the wrinkling
and pull-in failures. After a short calculation, we deduce that
the optimal value of $\Delta S$
\begin{equation}
\Delta S_{\mbox{\tiny \it opt}}^{\mbox{\tiny \it pull}}=\mu(\nu_
{\mbox{\tiny \it opt}}^{\mbox{\tiny \it pull}}-(\nu_{\mbox{\tiny
\it opt}}^{\mbox{\tiny \it pull}})^{-2}),\label{deltas}
\end{equation}
whereas the optimal value of the electric field is  given by
\begin{equation}
E_{\mbox{\tiny \it opt}}^{\mbox{\tiny \it pull}}=
\sqrt{ \frac{3 \mu}{\varepsilon}} \left ((1-\chi)^{-\frac{1}{6}}
(4-\chi)^{-\frac{1}{3}}\right ). \label{ee}
\end{equation}
\begin{figure}[th]
\begin{center}
\includegraphics[height=7.0 cm]{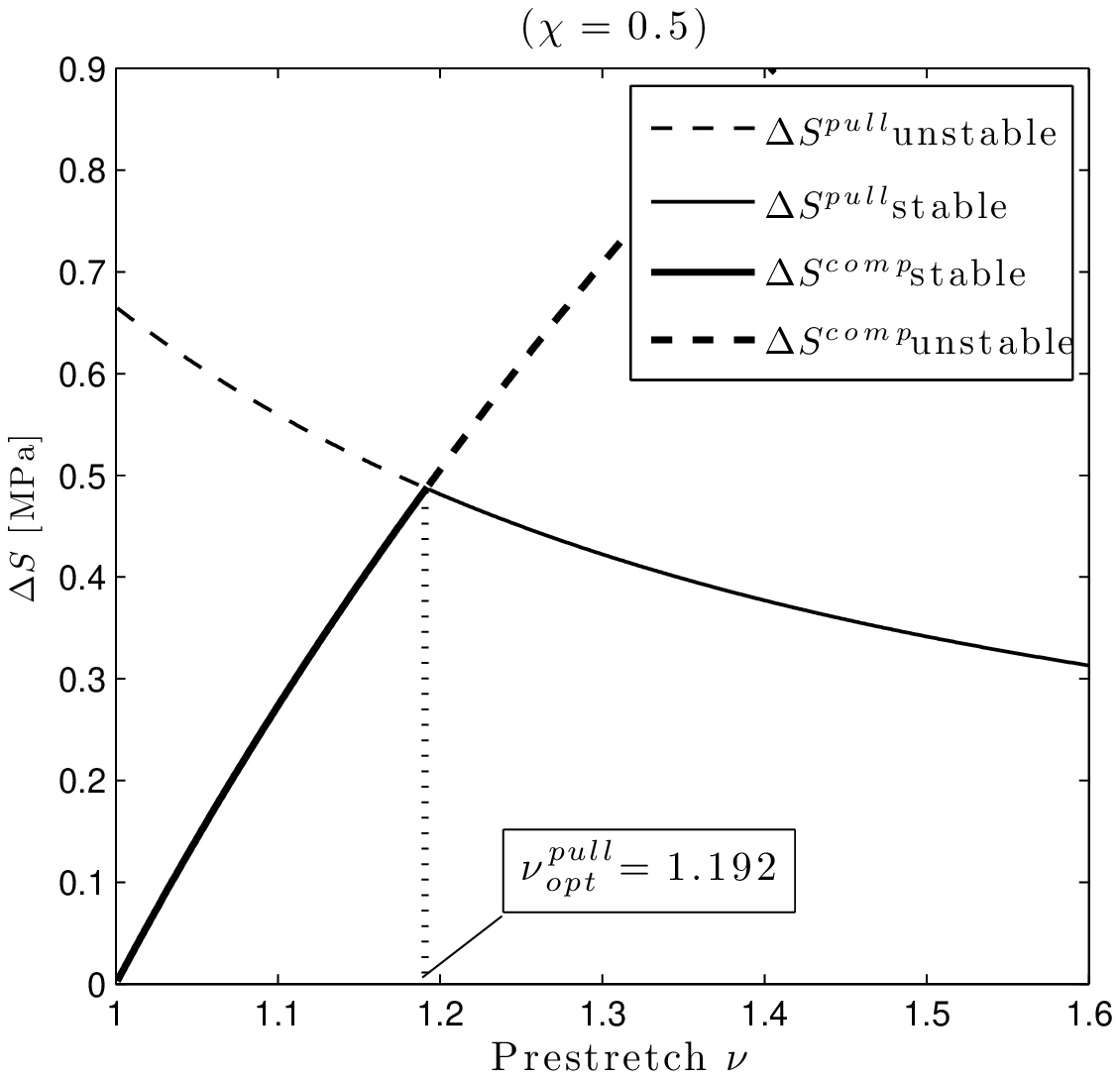}a)\\
\includegraphics[height=6.5 cm, width=8 cm]{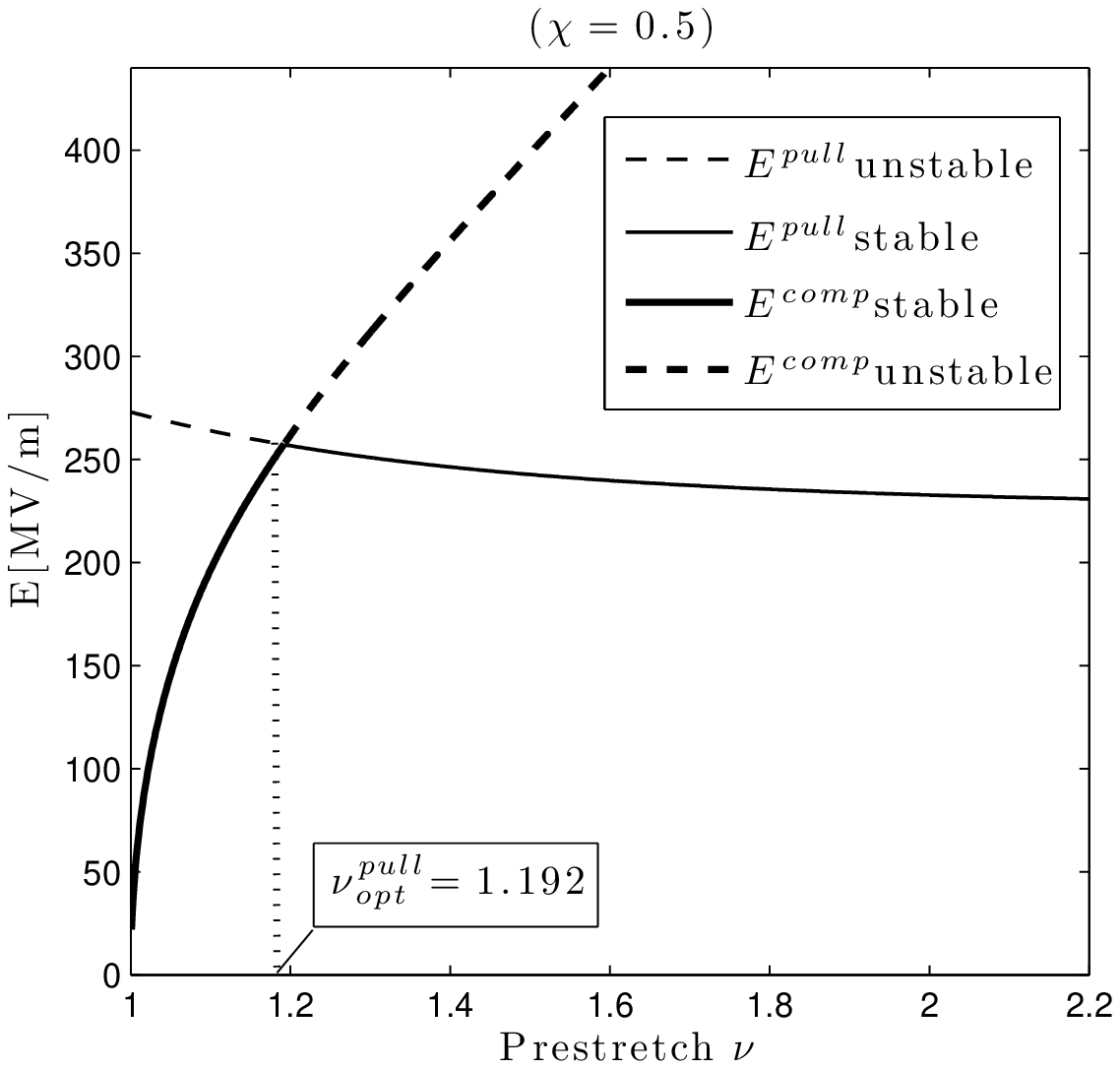}b)\\
\vspace{0.3 cm}\caption{\label{Optimvolt} Critical values of the activation
stress  $\Delta S$ a) and of the electric field $E$. Bold lines represent the
values corresponding to wrinkling instability, whereas light lines represent
the values corresponding to pull-in instability. Dashed lines correspond to
unstable equilibrium states. }
\end{center}
\end{figure}
It should be underlined, in conclusion, that all the activation parameters $\D
S, \kv$ and $\bE$ depend on the elastic modulus $\mu$, and that $\D S $ and
$\bE$ grow with the fraction $\chi$ of activated film. The resulting dependence
of $\nu^{\mbox{\it \tiny pull}}$ and $\nu^{\mbox{\it \tiny comp}}$ with respect
to the fraction of activated material is summarized in Fig.\ref{optimni}.

\newpage
\subsection{Charge Control}

In the case of charge control, it is interesting to observe that the pull-in
failure is not possible (this is in accordance with previous analyses
\cite{SB}). This is due to the fact that for any $\nu$ and $\chi$ the stress
$s$ is a decreasing function of $\hat\lambda$, whereas the stress $\hat{s}$ is
an increasing function of $\hat\lambda$. On the contrary the wrinkling failure
is possible and it is attained as soon as $s(\nu,\chi)=\hat s(\kq)=0$, that is
$$
\hat \lambda=\hat \lambda_{\mbox{\tiny \it opt}}^{\mbox{\tiny \it comp}}(\nu)=1+\frac{\nu-1}{\chi}
$$
which corresponds to a maximum load parameter
$$
\kappa_{\mbox{\tiny \it Q,max}}^{\mbox{\tiny \it
comp}}=\frac{1}{2}\left ((\hat \lambda_{\mbox{\tiny \it
opt}}^{\mbox{\tiny \it comp}})^{6}-1 \right ).
$$
Since the load parameter is an increasing function of $\nu$, it is
clear that in the case of charge control it is not possible to
find an optimal value of the prestretch.
\begin{figure}[th]\begin{center}
\includegraphics[height=7.1 cm]{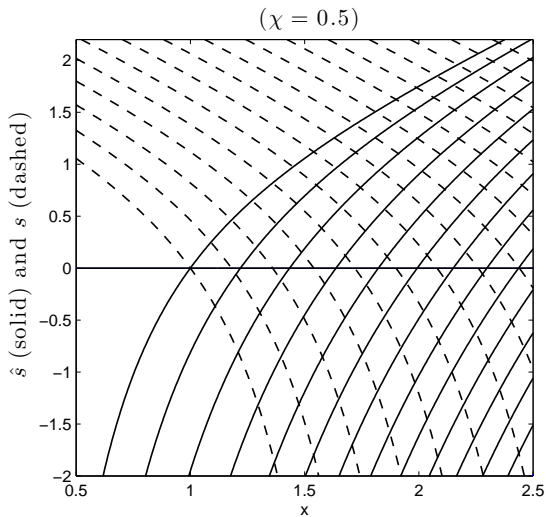}
\vspace{-0.4 cm}\caption{\label{Eq-charge} Stress strain curves in
the activated and non activated region in the case of assigned
charge: continuous lines represent the stress $\hat{s}_{1}(\hat
\lambda, \kq)$ in (\ref{Sc11}), of the activated region, whereas
dashed lines represent the stress $s_{1}(\hat \lambda, \nu)$ in
(\ref{Sa11}) in the non active region.  }
\end{center}
\end{figure}

\section{Comparison with experiments}\label{exp}

\begin{figure}[h!]
\begin{center}
\includegraphics[width=8cm, height=7 cm] {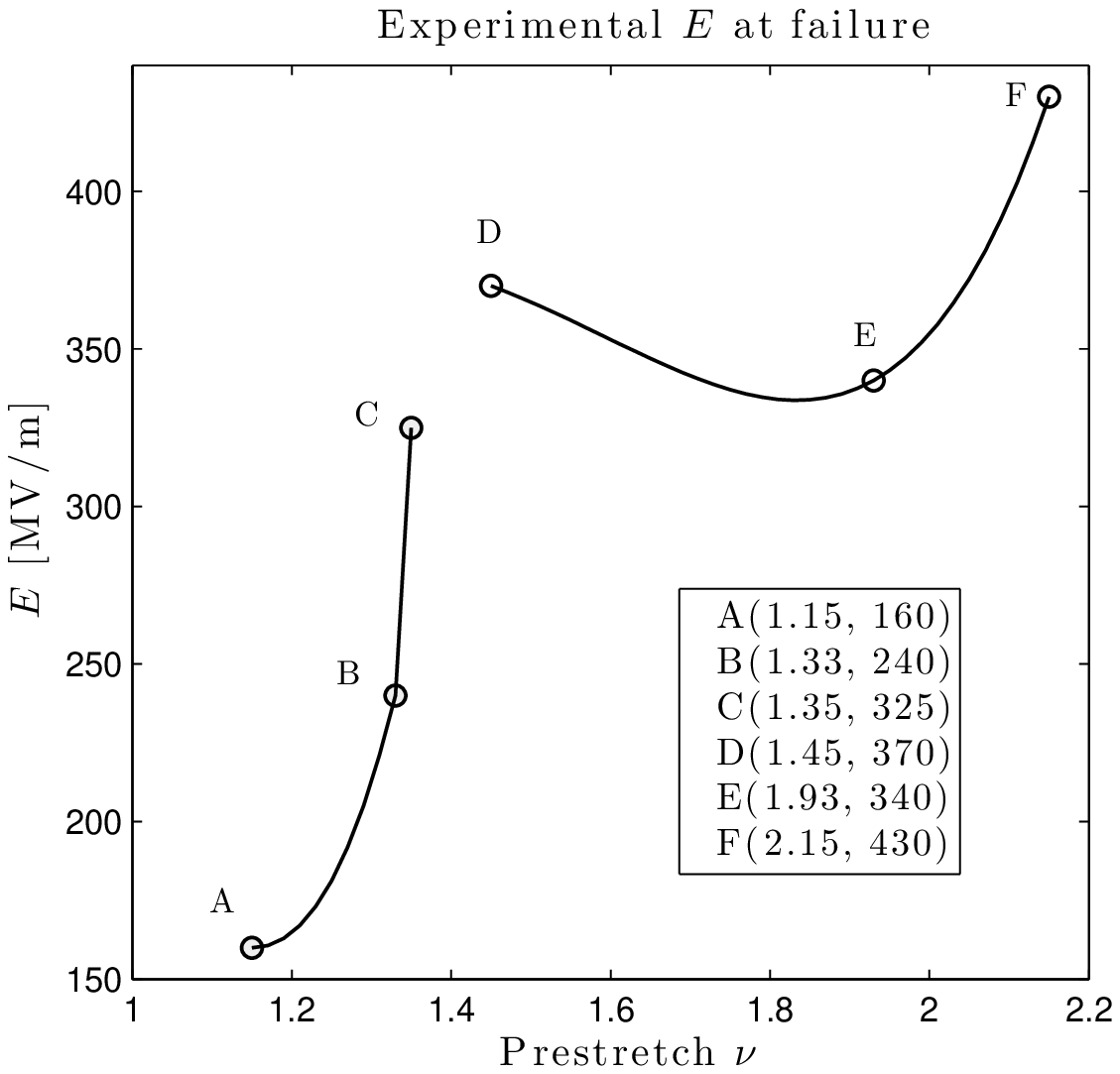}a) \\
\hspace{.2 cm}\includegraphics[width=7.8cm, height=7 cm]{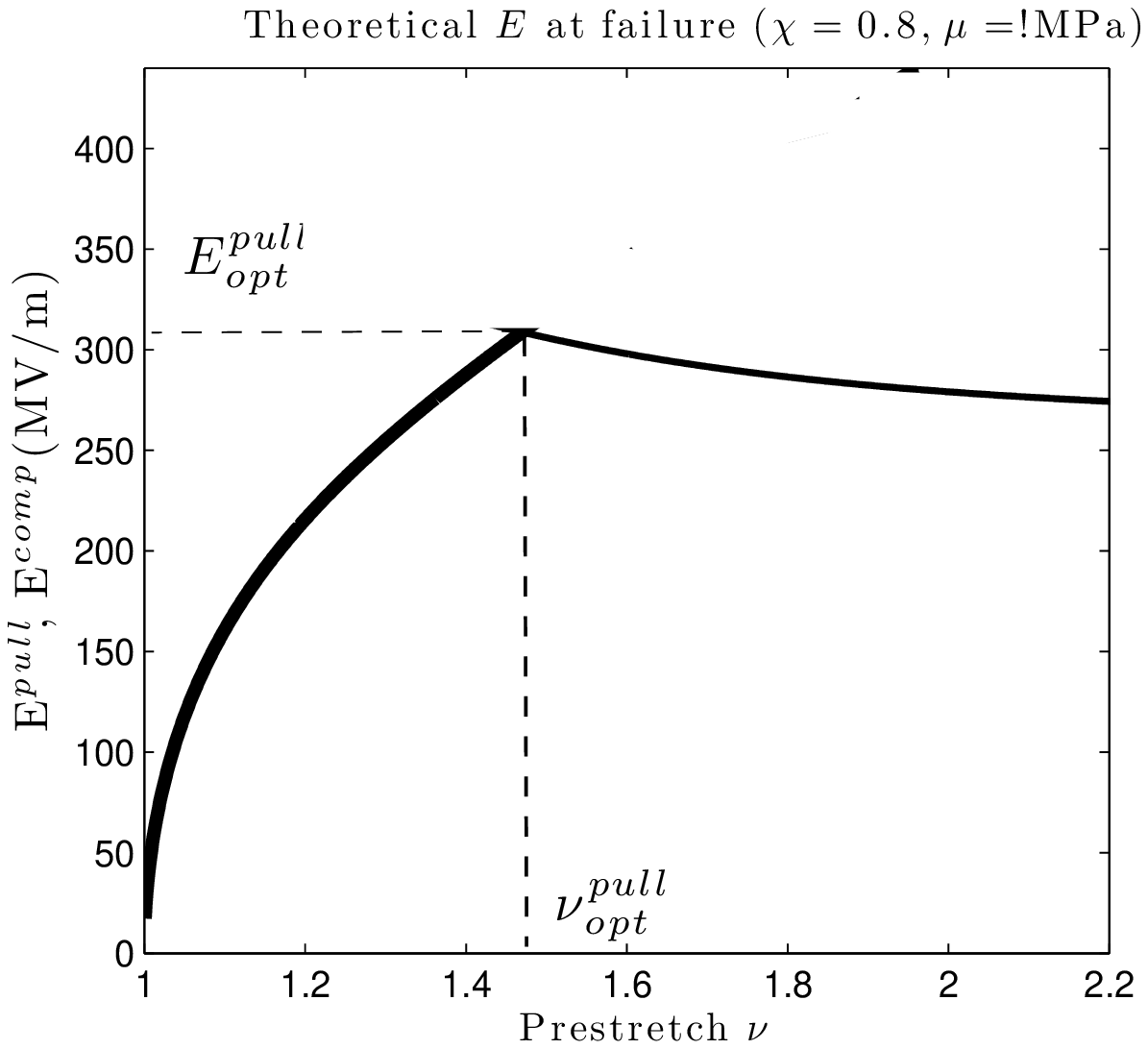}b)\\
\vspace{-0.1 cm}\caption{\label{Exper}  Failure value of the
electric field as a function of the prestretch: a) experiment
(deduced from \cite{KP}), b) theoretical.  }
\end{center}
\end{figure}

In this section we briefly compare the theoretical results of our
model with the experimental behavior of electroactive polymer
actuators \cite{KP}.

In Fig.\ref{Exper}$_{a}$ we reproduce the experimental behavior of
silicone rubber films (Nusil CF19-2186) as reported in \cite{KP},
where circles represent the values of the electric field at
failure for different assigned prestretches. In
Fig.\ref{Exper}$_b$ we reproduce the results of our model, where
we have considered an elastic modulus $\mu=1 MPa$, as suggested
by \cite{KP} for this material. It should be remarked that
the considered experiments refer to circular actuators, whereas we here
have focused attention on linear actuators. Nevertheless, the
comparison between experiments and theoretical results are in nice
quantitative agreement.

Quite interestingly, a careful analysis of the experimental
Fig.\ref{Exper}$_a$ in light of the theoretical
Fig.\ref{Exper}$_b$ shows the existence of different behaviors,
separated by a threshold prestretch which approximately
corresponds to $\nu_{\mbox{\tiny \it opt}}^{\mbox{\tiny \it
pull}}$. For values of $\nu$ lower than $\approx 1.4$
 both the experimental and the theoretical results show an
initial growth of the maximum applicable electric field with
growing prestretch. According to our model this may be due to the
initially positive effect of prestretch in avoiding the occurrence
of wrinkling. Such positive effect, both in experiments and in our
model, ceases to exist when the prestretch overcomes the critical
value of about $1.4$, since here the thickness reduction
anticipates the occurrence of pull-in failure. A finer description
of the film behavior for larger values of the prestretch certainly
requires a better description of the exact film geometry and of
the material response. To the price of renouncing to closed form
solutions (which are generally possible under very simple
assumptions such those embraced in this article), the methods we
have discussed can be numerically applied to more complex
geometries and material behaviors suitable for polymers undergoing
large deformations, such as Mooney-Rivlin, Gent, Arruda-Boyce and
others; some of these applications will be considered in a
forthcoming paper.

\section*{APPENDIX}
\setcounter{section}{0} \setcounter{equation}{0}
\renewcommand{\theequation}{A.\arabic{equation}}

In this Appendix a necessary stability condition is obtained for
the equilibrium configurations of a linear actuator undergoing
piecewise homogeneous deformations. We observe that the
homogeneity of the deformations in both the active and non active
regions, together with the constraint of assigned prestretch
$\nu$, essentially reduces the present equilibrium problem to a one-dimensional problem. Under conservative
hypotheses, the total energy $\scrE$ stored in the device at given
$\hat \lambda$ can be obtained by integration of the internal
working density along an arbitrary deformation path from the
undeformed configuration; since in this case the only non zero
stress component is $s$ we have
$$
\frac{\scrE(\hat\lambda)}{HL}=\left[\chi\int_1^{\hat\lambda}
\hat{s} (x)\,dx + (1-\chi) \int_1^{\lambda} s(x)\,dx\right].
$$
This function must be minimized under the compatibility constraint
$\nu=\chi \hat \lambda+(1-\chi)\lambda$, that gives
$$
\lambda=\lambda(\hat \lambda)=\frac{\nu-\chi \hat \lambda}{1-\chi}.
$$
Let then $\hat\lambda$ define a given configuration of the
system, and let $\hat\lambda+\zeta$ be a perturbed
configuration that is obtained by shifting the interface between
the active and non active regions. The perturbed energy is
$$
\frac{\scrE(\hat\lambda+\zeta)}{HL}=\chi\int_1^{\hat\lambda+\zeta}
\hat{s}(x)\,dx +(1-\chi)
\int_1^{\lambda-\zeta\frac{\chi}{1-\chi}}s(x)\,dx.
$$
Thus, a stable configuration $\hat\lambda$ must verify
$$
\scrE(\hat\lambda+\zeta)\geq\scrE(\hat
\lambda)\hspace{20pt}\text{for all}\hspace{5pt}\zeta.
$$
By considering the following expansion
$$\begin{array}{lll}
\scrE(\hat\lambda+\zeta) &=& \scrE(\hat\lambda) +
\zeta\chi\left[\hat{s}(\hat\lambda)-s(\lambda)\right]\vspace{0.2 cm}\\
&+& \displaystyle \chi\frac{\zeta^2}
{2}\left[\frac{d\hat{s}(\hat\lambda)}{d\hat\lambda}-\frac{d
s(\lambda(\hat\lambda))}{d\hat\lambda}\right]+o(\zeta^2),\end{array}
$$
the vanishing of the first variation gives the Euler Lagrange
equilibrium condition
$$
\hat{s}(\hat\lambda)=s(\lambda),
$$
whereas  the positivity of the second variation gives
$$
\frac{d\hat{s}(\hat\lambda)}{d\hat\lambda}\geq\frac{d
s(\lambda(\hat\lambda))}{d\hat\lambda},
$$
which is the necessary stability condition adopted in the
paper.


\begin{thebibliography}{99}
\bibitem{BL}
J. Block, D.G. LeGrand. Dielectric breakdown of polymer films,
\textit{J. Appl. Phys.} \textbf{40}, 288--293, 1969.
\bibitem{DO} R. Bustamante, A. Dorfman, R.W. Ogden. Nonlinear Electroelastostatics:
a variational framework. \textit{Z. angew. Math. Phys.},
\textbf{60}, 154--177, 2009.
\bibitem{CRK} F. Carpi, D. De Rossi, R- Kornbluh, R. Perline, and P. Sommer-Larsen Ed.
\newblock {\em Dielectric elastomers as electromechanical transducers}.
\newblock Elsevier,  2008.
\bibitem{CGS}
I. Carr, M.E. Gurtin, and M.Slemrod. Structured phase transition on a finite interval.
\newblock \textit{Arch. Rat. Mech. Anal.}, \textbf{86}, 317--351, 1984.
\bibitem{C} B.D. Coleman. Necking and Drawing in Polymeric Fibers
Under Tension, \textit{Arch. Rat. Mech. An.} \textbf{83}(2),
115--137, 1983.
\bibitem{DPS1} D. De Tommasi, G. Puglisi, G. Saccomandi. A micromechanics based model
for the {M}ullins effect. \textit{J. Rheology}, \textbf{50},
495--512, 2006.
\bibitem{DPS2} D. De Tommasi, G. Puglisi, G. Saccomandi. Localized versus Diffuse
Damage in Amorphous Materials. \textit{Phys. Rev. Lett.},
\textbf{100}, 085502, 2008.
\bibitem{FU} Y. B. Fu, A. B. Freidin.
Characterization and stability of two-phase piecewise-homogeneous
deformations. \textit{Proc. R. Soc. Lond.  }, \textbf{460},
3065--3094 , 2004.
\bibitem{HO} D. M. Haughton and R. W. Ogden. On the incremental equations in non-linear elasticity-Membrane theory. \textit{J. Mech. Phys. Solids}. \textbf{26},
93--110, 1978.
\bibitem{K} G. Kofod. The static actuation of dielectric elastomer actuators: how does pre-stretch improve actuation. \textit{J. Phys. D: Appl. Phys.},
\textbf{41}, 215405, 2008.
\bibitem{KP} R. Kornbluh, R. Perline, Q. Pei, S. Oh, J. Joseph. Ultrahigh strain response
of field-actuated elastimeric polymers. \textit{Multibody System
Dynamics}, \textbf{1}, 149--188, 1997.
\bibitem{Par}
G. W. Parker. Electric field outside a parallel plate capacitor.
\textit{Am. J. Phys.}, \textbf{70}, 502--507, 2002.
\bibitem{PGS}
L. Patrick, K. Gabor, and M. Silvain. Characterization of
dielectric elastomer actuators based on a hyperelastic film model.
\textit{Sensors and Actuators A}, \textbf{135}, 748--757, 2007.
\bibitem{PKP} R. Perline, R. Kornbluh, Q. Pei, and J. Joseph. High speed electrically actuated elastomers with strain greater that 100\%. \textit{Science},
\textbf{287}, 836--839, 2000.
\bibitem{PD} S. Plante, S. Dubowsky. Large-scale
Failure modes of dielectric elastomer actuators. \textit{Int. J.
Sol. Struct.}, \textbf{43}, 7727--51, 2006.
\bibitem{Pip} A.C. Pipkin. The relaxed energy density for isotropic elastic thin layers. \textit{IMA J. Appl. Math.}, \textbf{36}, 85--99, 1986.
\bibitem{GP} G. Puglisi. Damage localization and stability in electroactive polymers. \textit{Report of Mathematisches Forschungsinstitut Oberwolfach}, \textbf{10}, 2008.
\bibitem{SB} J.I. Seeger and B.E. Boser. Charge control of parallel-plate, electrostatic actuators and the tip-in instability. \textit{J. Microelectromechanical Syst.}, \textbf{12}, 2003.
\bibitem{Ste} D.J. Steigmann. Tension-field theory. \textit{Proc. R. Soc. Lond. A}, \textbf{429}, 141--173, 1990.
\bibitem{ZS} X. Zhao and Z. Suo. Method to analyze electromechanical stability of dielectric elastomers. \textit{J. Appl. Phys.}, \textbf{91}, 061921, 2007.
\bibitem{ZHS} X. Zhao, W. Hong and Z. Suo. Electromechanical hysteresis and coexistent states in dielectric elastomers. \textit{Phys. Rev. B}, \textbf{76}, 134113, 2007.
\end{thebibliography}
\end{document}